
%
\magnification=1200
%
%
\hsize=31pc
\vsize=55 truepc
\hfuzz=2pt
\vfuzz=4pt
\pretolerance=5000
\tolerance=5000
\parskip=0pt plus 1pt
\parindent=16pt
%
%
\font\fourteenrm=cmr10 scaled \magstep2
\font\fourteeni=cmmi10 scaled \magstep2
\font\fourteenbf=cmbx10 scaled \magstep2
\font\fourteenit=cmti10 scaled \magstep2
\font\fourteensy=cmsy10 scaled \magstep2
\font\large=cmbx10 scaled \magstep1
%
\font\sans=cmssbx10
%
%
\def\bss#1{\hbox{\sans #1}}
%
%
\font\bdi=cmmib10
%
%
\def\bi#1{\hbox{\bdi #1\/}}
%
%
\font\eightrm=cmr8
\font\eighti=cmmi8
\font\eightbf=cmbx8
\font\eightit=cmti8

\font\eightsy=cmsy8
\font\sixrm=cmr6
\font\sixi=cmmi6
\font\sixsy=cmsy6

\def\tenpoint{\def\rm{\fam0\tenrm}%
  \textfont0=\tenrm \scriptfont0=\sevenrm
                      \scriptscriptfont0=\fiverm
  \textfont1=\teni  \scriptfont1=\seveni
                      \scriptscriptfont1=\fivei
  \textfont2=\tensy \scriptfont2=\sevensy
                      \scriptscriptfont2=\fivesy
  \textfont3=\tenex   \scriptfont3=\tenex
                      \scriptscriptfont3=\tenex
  \textfont\itfam=\tenit  \def\it{\fam\itfam\tenit}%
  \textfont\slfam=\tensl  \def\sl{\fam\slfam\tensl}%
  \textfont\bffam=\tenbf  \scriptfont\bffam=\sevenbf
                            \scriptscriptfont\bffam=\fivebf
                            \def\bf{\fam\bffam\tenbf}%
  \normalbaselineskip=20 truept
  \setbox\strutbox=\hbox{\vrule height14pt depth6pt
width0pt}%
  \let\sc=\eightrm \normalbaselines\rm}
\def\eightpoint{\def\rm{\fam0\eightrm}%
  \textfont0=\eightrm \scriptfont0=\sixrm
                      \scriptscriptfont0=\fiverm
  \textfont1=\eighti  \scriptfont1=\sixi
                      \scriptscriptfont1=\fivei
  \textfont2=\eightsy \scriptfont2=\sixsy
                      \scriptscriptfont2=\fivesy
  \textfont3=\tenex   \scriptfont3=\tenex
                      \scriptscriptfont3=\tenex
  \textfont\itfam=\eightit  \def\it{\fam\itfam\eightit}%
  \textfont\bffam=\eightbf  \def\bf{\fam\bffam\eightbf}%
  \normalbaselineskip=16 truept
  \setbox\strutbox=\hbox{\vrule height11pt depth5pt width0pt}}
\def\fourteenpoint{\def\rm{\fam0\fourteenrm}%
  \textfont0=\fourteenrm \scriptfont0=\tenrm
                      \scriptscriptfont0=\eightrm
  \textfont1=\fourteeni  \scriptfont1=\teni
                      \scriptscriptfont1=\eighti
  \textfont2=\fourteensy \scriptfont2=\tensy
                      \scriptscriptfont2=\eightsy
  \textfont3=\tenex   \scriptfont3=\tenex
                      \scriptscriptfont3=\tenex
  \textfont\itfam=\fourteenit  \def\it{\fam\itfam\fourteenit}%
  \textfont\bffam=\fourteenbf  \scriptfont\bffam=\tenbf
                             \scriptscriptfont\bffam=\eightbf
                             \def\bf{\fam\bffam\fourteenbf}%
  \normalbaselineskip=24 truept
  \setbox\strutbox=\hbox{\vrule height17pt depth7pt width0pt}%
  \let\sc=\tenrm \normalbaselines\rm}

\def\today{\number\day\ \ifcase\month\or
  January\or February\or March\or April\or May\or June\or
  July\or August\or September\or October\or November\or
December\fi
  \space \number\year}
%
%
\newcount\secno      
\newcount\subno      
\newcount\subsubno   
\newcount\appno      
\newcount\tableno    
\newcount\figureno   
\normalbaselineskip=20 truept
\baselineskip=20 truept
%
%
\def\title#1
   {\vglue1truein
   {\baselineskip=24 truept
    \pretolerance=10000
    \raggedright
    \noindent \fourteenpoint\bf #1\par}
    \vskip1truein minus36pt}
%
%
\def\author#1
  {{\pretolerance=10000
    \raggedright
    \noindent {\large #1}\par}}
%
%
\def\address#1
   {\bigskip
    \noindent \rm #1\par}
%
%
\def\shorttitle#1
   {\vfill
    \noindent \rm Short title: {\sl #1}\par
    \medskip}
%
%
\def\pacs#1
   {\noindent \rm PACS number(s): #1\par
    \medskip}
%
%
\def\jnl#1
   {\noindent \rm Submitted to: {\sl #1}\par
    \medskip}
%
%
\def\date
   {\noindent Date: \today\par
    \medskip}
%
%
\def\beginabstract
   {\vfill\eject
    \noindent {\bf Abstract. }\rm}
%
%
\def\keyword#1
   {\bigskip
    \noindent {\bf Keyword abstract: }\rm#1}
%
%
\def\endabstract
   {\par
    \vfill\eject}
%
%
%

%
%
\def\entry#1#2#3
   {\noindent
    \hangindent=20pt
    \hangafter=1
    \hbox to20pt{#1 \hss}#2\hfill #3\par}
%
%
\def\subentry#1#2#3
   {\noindent
    \hangindent=40pt
    \hangafter=1
    \hskip20pt\hbox to20pt{#1 \hss}#2\hfill #3\par}
%
%
\def\section#1
   {\vskip0pt plus.1\vsize\penalty-250
    \vskip0pt plus-.1\vsize\vskip24pt plus12pt minus6pt
    \subno=0 \subsubno=0
    \global\advance\secno by 1
    \noindent {\bf \the\secno. #1\par}
    \bigskip
    \noindent}
%
%
\def\subsection#1
   {\vskip-\lastskip
    \vskip24pt plus12pt minus6pt
    \bigbreak
    \global\advance\subno by 1
    \subsubno=0
    \noindent {\sl \the\secno.\the\subno. #1\par}
    \nobreak
    \medskip
    \noindent}
%
%
\def\subsubsection#1
   {\vskip-\lastskip
    \vskip20pt plus6pt minus6pt
    \bigbreak
    \global\advance\subsubno by 1
    \noindent {\sl \the\secno.\the\subno.\the\subsubno. #1}\null. }
%
%
\def\appendix#1
   {\vskip0pt plus.1\vsize\penalty-250
    \vskip0pt plus-.1\vsize\vskip24pt plus12pt minus6pt
    \subno=0
    \global\advance\appno by 1
    \noindent {\bf Appendix \the\appno. #1\par}
    \bigskip
    \noindent}
%
%
\def\subappendix#1
   {\vskip-\lastskip
    \vskip36pt plus12pt minus12pt
    \bigbreak
    \global\advance\subno by 1
    \noindent {\sl \the\appno.\the\subno. #1\par}
    \nobreak
    \medskip
    \noindent}
%
%

%
%

%
%
\def\tabcaption#1
   {\global\advance\tableno by 1
    \noindent {\bf Table \the\tableno.} \rm#1\par
    \bigskip}
%
%
\def\figures
   {\vfill\eject
    \noindent {\bf Figure captions\par}
    \bigskip}
%
%
\def\figcaption#1
   {\global\advance\figureno by 1
    \noindent {\bf Figure \the\figureno.} \rm#1\par
    \bigskip}
%
%
\def\references
     {\vfill\eject
     {\noindent \bf References\par}
      \parindent=0pt
      \bigskip}
%
%
\def\refjl#1#2#3#4
   {\hangindent=16pt
    \hangafter=1
    \rm #1
   {\frenchspacing\sl #2
    \bf #3}
    #4\par}
%
%
\def\refbk#1#2#3
   {\hangindent=16pt
    \hangafter=1
    \rm #1
   {\frenchspacing\sl #2}
    #3\par}
%
%
\def\numrefjl#1#2#3#4#5
   {\parindent=40pt
    \hang
    \noindent
    \rm {\hbox to 30truept{\hss #1\quad}}#2
   {\frenchspacing\sl #3\/
    \bf #4}
    #5\par\parindent=16pt}
%
%
\def\numrefbk#1#2#3#4
   {\parindent=40pt
    \hang
    \noindent
    \rm {\hbox to 30truept{\hss #1\quad}}#2
   {\frenchspacing\sl #3\/}
    #4\par\parindent=16pt}
%
%

%
%
\def\frac#1#2{{#1 \over #2}}
%
%

%
%
\def\d{{\rm d}}
%
%
\def\e{{\rm e}}
%
%
\def\i{\ifmmode{\rm i}\else\char"10\fi}
%
%

%
%

%
%

%
%

%
%
\catcode`\@=11
%
%
\def\ind{\hbox to 5pc{}}
%
%
\def\eq(#1){\hfill\llap{(#1)}}
%
%

%
%
\def\deqn#1{\displ@y\halign{\hbox to \displaywidth
    {$\@lign\displaystyle##\hfil$}\crcr #1\crcr}}
%
%
\def\indeqn#1{\displ@y\halign{\hbox to \displaywidth
    {$\ind\@lign\displaystyle##\hfil$}\crcr #1\crcr}}
%
%
\def\indalign#1{\displ@y \tabskip=0pt
  \halign to\displaywidth{\ind$\@lign\displaystyle{##}$\tabskip=0pt
    &$\@lign\displaystyle{{}##}$\hfill\tabskip=\centering
    &\llap{$\@lign##$}\tabskip=0pt\crcr
    #1\crcr}}
\catcode`\@=12
%
%



%
%



%
%

%
%
\def\\{\hfill\break}
\def\avd#1{\overline{#1}}
\def\avt#1{\langle#1\rangle}

\def\bbox#1{\bi{#1}}
\def\s{\sigma}
\def\Pcal{{\cal{ P}}}

\title{Lack of self-average in weakly disordered one dimensional systems}

\author{A. Crisanti\dag, G. Paladin\ddag, M. Serva\S\ and A. Vulpiani\dag}

\address{\dag\ Dipartimento di Fisica, Universit\`a `La Sapienza' \\
               P.le Moro 5, I-00185, Roma, Italy}

\address{\ddag\ Dipartimento di Fisica,  Universit\`a dell'Aquila \\
                I-67100 Coppito, L'Aquila, Italy}

\address{\S\ Dipartimento di Matematica,  Universit\`a dell'Aquila \\
             I-67100 Coppito, L'Aquila, Italy}


\beginabstract
We introduce a one dimensional disordered Ising model which
at zero temperature is characterized by a non-trivial,
non-self-averaging, overlap probability distribution
when the impurity concentration vanishes in the thermodynamic limit.
The form of the distribution can be calculated analytically for any
realization of disorder.
For non-zero impurity concentration the distribution becomes
a self-averaging delta function centered on a value
which can be estimated by the product of appropriate transfer matrices.
\endabstract

\section{Introduction}
Disordered systems attracted much work in the last years. One of the main
features of these systems is the large number of locally stable states.
As a consequence, a natural characterization of the equilibrium states is
through the distribution of their mutual distance, that is through the
probability distribution $P(q)$ of the overlap $q$ between pairs of states
[1].
In general, the organization of the  equilibrium states depends on the
realization of disorder, so that $P(q)$ depends on the disorder, even after
the thermodynamic limit has been taken.
 For instance, this lost of self-averaging is
observed  in the Sherrington-Kirkpatrick model of spin glasses [2]
where it can be related to a symmetry breaking in the replica space [1].
A big effort has been devoted to the search of simpler systems
which share the main features of the Sherrington-Kirkpatrick model.
The most celebrated among them is the random energy model [3]
now used in very different contexts such as biological
 evolution, neural networks, polymers, and so on.
On the other hand, it is commonly assumed that one dimensional systems lack
for such a rich behaviour. In this paper we introduce a disordered Ising
model which, in spite of being a one dimensional model with short range
interactions, exhibits at zero temperature a non-trivial, non-self-averaging,
$P(q)$ if the impurity concentration vanishes in the thermodynamic limit.
This latter condition is crucial, since if the disorder concentration does
not vanish in the thermodynamic limit, $P(q)$ becomes a self-averaging delta
function, even at zero temperature. This is the main result of the paper
which allows one  to get a deeper understanding of the mechanisms underlying
the appearance of non-trivial ergodicity breaking phenomena in disordered
systems.

The paper is organized as follows.
In Sect.~2, we introduce the model and the thermodynamic
quantities of interest.
In Sect.~3, we study the model in the weakly disorder limit,
where it is possible to calculate the form of the
overlap distribution $P(q)$ for any disorder realization.
In Sect.~4, the model is analyzed for non-zero impurity concentration
where $P(q)$ is shown to be a self-averaging delta function.
The value of the overlap where the delta function is centered can be obtained
by means of random transfer matrices.

In Appendix 1, we discuss an extension of the Landau theorem on the absence of
phase transitions in one dimension to the overlap probability distribution.
In Appendix 2, we give a brief description of the numerical algorithm
used in Sect.~4 to compute thermodynamical quantities via product of
transfer matrices [4].

\section{The system}
One of the main reasons for the existence of
a large number of equilibrium states in statistical systems is
the presence of ``frustration''.
In general, a disordered system cannot  satisfy all the
constraints originated from the competing effects of disorder and
interactions. We introduce a one dimensional model which captures
the main features of disordered systems but, at the same time, is simple
enough for an analytical study.
The model is a random field Ising chain of $N$ spins described by the
Hamiltonian
$$
H=-\sum_{i=1}^N \bigl[ \s_i \, \s_{i+1}+h_i \, \s_i \bigr]
\eqno(2.1)$$
with
$$
h_i= \eta_i -\eta_{i+1}
\eqno(2.2)$$
where the $\eta_i$ are random independent variables which take the value
$+1$ and $-1$ with probability $p$ and $(1-p)$, respectively. As a
consequence, the magnetic field on a site assumes the value $h_i=\pm 2$
with probability $p \, (1-p)$, and  $h_i=0$ with probability
$p^2 + (1-p)^2$. By changing the value of $p$, we can modulate the disorder in
the model. The pure ferromagnetic system is obtained for $p=1$ or $0$,
yielding $h_i=0$ $\forall \, i$. We can also consider a pure system
with a finite number $I$ of impurities, so that the impurity concentration
vanishes in the thermodynamic limit, {\it weakly disordered system}.
The integer number $I$ is a random variable  distributed according to
$P(I)=c^I \exp(-c)/I!$ where $c=\sum_I P(I) \, I $  is its mean value.
The weakly disorder limit is achieved by taking  $\eta_i=1$
with probability $p=1-c/N$, and hence $p\to 1$ when the number of spins
$N\to\infty$.

For any value of $p$ the local magnetic field has a certain degree of
spatial correlation, all field realizations are made by a sequence of strings
$$
 2 \, , \ 0 \cdots \ 0 \, , -2  \, , \ 0 \cdots \ 0 \ 2
$$
where the number of zeros between two successive
$2$ and $-2$ or $-2$ and $2$ is a random non-negative variable.

The overlap between two spin configurations `$\alpha$' and `$\beta$'
for the same disorder realization is defined as
$$
q_{\alpha \beta}= {1\over N} \sum_{i=1}^N
    \sigma_i^{\alpha}\,\sigma_i^{\beta}.
$$
This can be usefully computed by introducing two identical replicas of the
system. The equilibrium probability distribution of $q_{\alpha\beta}$
is, therefore, given by
$$
P(q;N)= Z_N^{-2}\,
 \sum_{\sigma^{1}} \sum_{\sigma^{2}}\,
 \delta (q- q_{12})\,  \prod_{\alpha=1,2} \exp(-\beta\, H^{\alpha})
\eqno(2.3)
$$
where $H^1$ and $H^2$ are the Hamiltonians (2.1) of the two replicas and
$\beta=1/T$. The
normalization factor $Z_N$ is the single replica partition function, which
can be defined in terms of the product of random transfer matrices as
$$
Z_N= \sum_{\sigma_i=\pm1} \prod_{i=1}^N \exp\beta
[\sigma_i \sigma_{i+1} \, + \eta_{i+1} \, (\sigma_{i+1} -\sigma_i)]
={\rm Tr} \, \prod_{i=1}^N \bss{A}_i
\eqno(2.4)$$
where, for this model,
$$\eqalign{
           \bss{A}_i&=\pmatrix{ e^{\beta} & e^{\beta}\cr
                                e^{-3\beta} & e^{\beta} }
                 \qquad \hbox{\rm with probability}\ p \cr
                    & \cr
           \bss{A}_i&=\pmatrix{ e^{\beta} & e^{-3\beta}\cr
                                e^{\beta} & e^{\beta} }
                 \qquad \hbox{\rm with probability}\ 1-p. \cr
          }
\eqno(2.5)$$
In general $P(q;N)$ depends on the particular sequence of random fields, even
in the thermodynamic  limit $N\to\infty$.
However, one can show that in one dimensional systems with short range
interactions for any finite temperature $P(q;N)$ is self-averaging, and hence
$P(q;N)$ converges towards a well defined function $P(q)$ as
$N\to\infty$.
This is no more true at zero temperature, where
the Landau  theorem on the absence of a phase transitions
does not hold, and one could  have an ergodicity breaking
[see Appendix 1].
The zero temperature limit of this model, with $p=1/2$, was studied
numerically in Ref.~5 by extracting the diverging part in (2.5), i.e. by
writing $\bss{A}_i=e^{\beta} \, \bss{R}_i$ with
$$
\bss{R}_i=\pmatrix{ 1 & 1 \cr
                  0 & 1}
\qquad \hbox{\rm or} \qquad
 \pmatrix{ 1 & 0 \cr
           1 & 1}
\eqno(2.6)$$
and using the matrices $\bss{R}_i$ in the product (2.4).
The non-zero value of the zero temperature entropy of the model found in
Ref.~5 indicates that frustration plays an important role.

\section{Ergodicity breaking in the weakly disordered system}
The weakly disordered system has a great theoretical relevance, because
it exhibits a non-trivial ergodicity breaking at zero temperature.
To our knowledge, this is the first disordered one dimensional model
where the overlap probability distribution $P(q)$ is a
non-self-averaging smooth function of $q$.

It has to be noted that
the pure system, with $h_i=0$ $\forall i$, also exhibits an ergodicity
breaking. The overlap between two equilibrium configurations
should be zero for any finite temperature, so that $P(q)=\delta(q)$.
On the other hand, at zero temperature the overlap is
$q=\pm1$ and $P(q)=[ \delta(q-1)+\delta(q+1)]/2$.
This is a quite trivial ergodicity breaking due to the
 zero temperature phase transition in the Ising model with zero field.
It simply reflects the breaking of the ``up-down'' symmetry of the model.

The weakly disordered systems is defined by assuming $\eta_i=1$ with
probability  $p=1-c/N$, where $N$ is the number of spins. As a
consequence, there is a finite average number $c$ of impurities in the
thermodynamic limit,  though $p=1$.

Any field realization of the weakly disordered system obeys some
simple rules. A sequence $\{ \eta_{k-1}=1, \ \eta_k=-1, \ \eta_{k+1}=1\}$
has probability $p^2 \, (1-p) \sim c/N$  while
a sequence $\{\eta_{k-1}=1, \ \eta_k=-1, \ \eta_{k+1}=-1\}$
has probability $p(1-p)^2 \sim (c/N)^2$, and the latter event can be
neglected in the limit of large $N$. It follows
that if $h_i=2$, then $h_{i+1}=-2$. Therefore in the thermodynamic limit,
a field configuration is given, with probability one, by
``islands'' of zero $h_i$ of arbitrary length
separated by $I$ interfaces made of the pair of random fields
$h_{j}=2$ and $h_{j+1}=-2$.

Within this picture, we can determine the overlap probability distribution
by combinatorial considerations.
Let us consider the case of $I=1$ impurity, that is particularly simple
since there is only one field interface.
Assuming periodic boundary conditions,   the interface can be moved
to the boundary, so we can take $h_1=-2$, $h_N=2$ and $h_i=0$ for
$i=2,\ldots N-1$.  The system  is thus equivalent to
a pure Ising model without external field and with fixed  boundary conditions
$\s_1=-1$ and $\s_N=+1$. At zero temperature, such a system is frustrated
because the spins  should be alligned with $\s_1$ on one side,
and with $\s_N$ on the other one. Consequently, there is a spin flip
in an arbitrary point $1<L<N$. The ground state is then made of spin
configurations with the first $L$ spins down and the remaining $N-L$
spins up. The overlap between two of such spin
configurations with spin-flip points $L$ and $L'$ is
$$q(L,L')=1- 2\,{|L-L'|\over N}. \eqno(3.1)$$
The probability distribution is obtained by counting how many times
the overlap is equal to $q$, i.e.
$$
P_1(q)={1\over N^2} \sum_{L=1}^N \, \sum_{L'=1}^N \delta(q-q(L,L'))
\eqno(3.2)$$
since all values of $L$ and $L'$ are equiprobable.
In the limit of large $N$ the sum can be estimated,
by defining the new variables $\ell=L/N$ and $x=|L-L'|/N$, as
$$
\int_0^1 \d\ell \, \int _0^\ell \d x \ \delta(q-1+2x).
\eqno(3.3)$$
The integration can be easily performed and yields
$$
P_1(q)=   \int_0^1 \d\ell \ \Theta(q-1+2\ell)={1+q\over 2}
\eqno(3.4)$$
It is worth stressing that $P_1(q)$ is
the same for all disorder configurations, due to the cyclic property of the
trace, though it is not a delta function. The
average overlap $\avd{q}=\int P_1(q) \, q \, \d q=1/3$
is different from the most probable value $q_{\rm mp}=1$

Consider now the case of $I = 2$ impurities, which is qualitatively different
from $I=1$. Each disorder realization has two islands of zero fields.
Assuming periodic boundary conditions, one interface can still be moved
to the boundaries, i.e. $h_1=-2$, $h_N=2$.  The other must be placed on
an arbitrary site $j=x \, N$, i.e. $h_j=2$, $ h_{j+1}=-2$.
In the limit of large $N$, the systems with $I=1$ and $I=2$ have the
same energy but different entropic factors since for $I=2$ there are $N$
allowed positions $j$ for the second interface. A ground state configuration
is individuated by the position of the interface, i.e. by the particular
value of $x$ in the interval $[0,1]$.

We can regard the system with $I=2$ impurities as the superposition of two
systems with $I=1$ impurity, made of $x \,  N$ and of $(1-x) \,  N$ spins,
respectively. The overlap between two ground states of the same disorder
realization $x$ is then given by the weighted sum of the  overlaps
of the two systems with $I=1$,
$$
q(x)= x \, q_1 + (1-x) \, q_2
\eqno(3.5)$$
where the $q_i$ are distributed according to
$$
P_1(q_i)={1+q_i\over 2}
\eqno(3.6)$$
At difference with the case $I=1$, for $I=2$ the overlap depends
on the particular disorder realization, implying
that the overlap probability distribution $P_2(q)$ is not self-averaging.
For any value of $x$, $P_2(q)$ is given by
$$
P_2(q;x)=\int_{-1}^1 \d q_1 \, \int_{-1}^1 \d q_2 \, P_1(q_1)
 \, P_1(q_2)\, \delta (q-q(x)).
\eqno(3.7)$$
By performing the first integral, we get
$$
P_2(q;x)={1\over x^2}
 \int_{a}^b \d q_2 \
 \left[ \, (x+q) \, + \, q_2 \, (q-1+2\, x) \, - \, q_2^2 \, (1-x) \,
 \right]
\eqno(3.8)$$
where
$$
a= \max \,  \left(-1 \, , \ {q-x \over 1-x} \right), \qquad
b= \min \, \left( 1 \, , \ {q+x \over 1-x} \right)
\eqno(3.9)$$
The overlap probability (3.8) is shown in Fig. 1 for different positions of
the second impurity ($x=0.1$, $x=0.2$ and $x=0.5$). For comparison,
we also report the average overlap probability distribution $\avd{P_2}(q)$.

For $I>2$ impurities, it is immediate to repeat the above argument.
Thus, the
overlap probability distribution of the system with $I$ field interfaces
 located on the sites
$x_1 \, N,\ldots, x_{I-1} \,  N,\, N$ is given by
$$
P_{I}(q;x_1,x_2,\ldots, x_{I-1})=\int_{-1}^1 \d q_1 \cdots
 \int_{-1}^1 \d q_{I}   \
\delta\left( q-\sum_{j=1}^I \,  x_j q_j \right) \,
 \  \prod_{j=1}^{I}  P_1(q_j)
\eqno(3.10)$$
with $x_{I}=1-\sum_1^{I-1} x_j$.
This expression takes a simple form in the Fourier space.
Let us define the Fourier transform of the overlap probability distribution
$$
P(\omega)=\int_{-\infty}^{\infty} \e^{\i \omega q} P(q) \d q
\eqno(3.11)$$
where $P(q)=0$ for $|q|>1$.
 In the case $I=1$, one can easily see that
$$
P_1(\omega)=\int_{-1}^1 \, {(1+q)\over 2}\, \e^{\i \omega q}\,  dq
 = \left( 1-\i {\partial \over \partial \omega} \right)\,
         {\sin(\omega)\over \omega}
\eqno(3.12)$$
Inserting into (3.10) the integral representation of the delta function,
we have for $I>1$ impurities we have
$$\eqalignno{
P_{I}(q;x_1,x_2,\ldots, x_{I-1})&=\int_{-\infty}^{\infty}
  {\d \omega\over 2\pi}\, \e^{-\i \omega q}
  \int_{-1}^1 \d q_1 \cdots  \int_{-1}^1 \d q_{I}
 \,  \prod_{i=1}^{I} P_1(q_i)\, \e^{\i \omega x_i q_i} \cr
    &=\int_{-\infty}^{\infty}
  {\d \omega\over 2\pi}\, \e^{-\i \omega q}
   \, \prod_{i=1}^{I} P_1(\omega \, x_i).
&(3.13)}$$
Therefore, the Fourier transform of the overlap probability distribution
is given by the convolution
$$
P_{I}(\omega;x_1,x_2,\ldots, x_{I-1})=
 \prod_{i=1}^{I} P_1(\omega x_i) \qquad
      \hbox{\rm with}\ \sum_{i=1}^{I} x_i=1
\eqno(3.14)$$
where
$$
P_1(\omega x_i)=
 {\sin (\omega\,  x_i)\over \omega \, x_i}- \i \,
 {\cos (\omega\,  x_i)\over \omega \, x_i}+ \i \,
 {\sin (\omega\,  x_i)\over (\omega \, x_i)^2}
\eqno(3.15)$$

For any number $I>1$ of impurities,
the overlap probability distribution depends on the
particular disorder configuration individuated by the
position of the $I-1$ field interfaces, i.e by the
  sequence  $\{x_1, \, x_2, \, \cdots \,  x_{I-1} \}$.
 The disorder average is, therefore, obtained by averaging over the
probability of the sequence $\Pcal(\{x_1,x_2,\ldots x_{I-1}\})$.
In  our model  the $x_i$ are independent random variables
uniformly distributed in the interval $[0,1]$, so that
$$
\Pcal(\{x_1, \, x_2, \, \cdots \,  x_{I-1} \}) =\prod_{i=1}^{I-1}
       {\cal P}(x_i)
\eqno(3.16)$$
where ${\cal P}(x)=1$ for $x \in [0,1]$ and ${\cal P}(x)=0$ otherwise.

Since the $x_i$ are independent, in the limit of large $I$
the average overlap probability distribution tends to the most probable value
$\widetilde{P}_{c}(\omega)$ obtained for $x_i=1/I$,
$$
\widetilde{P}_{c}(\omega)=\left[P_1(\omega)\right]^{c}
\eqno(3.17)$$
since for large $I$ the fluctuations about the mean value $c$ are negligible.
For the central limit theorem, $\widetilde{P}_{c}$ is a gaussian with mean
value $\avd{q}=1/3$ and variance $\avd{(q-1/3)^2} = 2/(9c)$.
 The gaussian form very quickly becomes  a good approximation,
 as shown in  Fig. 2 where
 $\widetilde{P}_{3}(q)$ (the most probable overlap probability
with impurities at $x_1=1/3$ and
 $x_2=2/3$) as well as  the average $\avd{P_3}(q)$ are very close to
to the gaussian with mean value $1/3$ and variance $2/27$.

The limit $c \to \infty$, but $c/N\to 0$,  corresponds to vanishing
impurity concentration  with $p \to 1$. As a consequence, at
zero temperature the overlap probability distribution is
$P(q)=\delta(q-1/3)$ for $p\to 1$,  while it is a double delta
$P(q)=[\delta(q-1)+\delta(q+1)]/2$ for
$p=1$, i.e. the pure system.

The weakly disordered system exhibits a non-trivial new phase between these
two extreme behaviours.
For finite temperature, the overlap vanishes both in the pure and in the weakly
disordered system leading to $P(q)=\delta(q)$.

\section{The disordered system}
In this section we discuss the system with non-zero impurity
concentration, i.e. with $\lim_{N \to \infty} c/N>0$ and $p\neq 1$ or $0$.
Unlike
the weakly disordered system, in this case we cannot perform simple
analytical calculations. However, we can employ the transfer matrix approach
for a numerical calculation [4].

The study of (2.3) is equivalent to that of [5,6]
   $${\cal N}_{N}(\omega)=
        \sum_{\sigma^{1}}\sum_{\sigma^{2}}\,
        \e^{N\,\omega\,q_{12}}\,
        \prod_{\alpha=1,2}\, \e^{-\beta H^{\alpha}}
   \eqno(4.1)$$
which can be regarded as the partition function
of a system made of two interacting replicas, with Hamiltonian
$H^1+H^2+\omega N q_{12}$. The term $\omega N$ is the macroscopic
coupling between the two replicas. Only if $\omega=0$, or
if the overlap $q_{12}=0$,  ${\cal N}_{N}(\omega)$  becomes the square
of the usual partition function (2.4).

The advantage of (4.1) stems from the possibility to
write it as a product of random transfer matrices [4,5]. The maximum Lyapunov
exponent $\Gamma(\omega)$ of the product yields the
average of the logarithm of ${\cal N}_N(\omega)$,
   $${\cal N}_N(\omega)\propto \exp[N\,\Gamma(\omega)]
    \qquad\hbox{\rm for}\ N \gg 1.
    \eqno(4.2)$$
There is no average over the disorder in the l.h.s. of (4.2)
since the Oseledec theorem [7] ensures that the Lyapunov exponent is a
non-random quantity for $N \to \infty$, i.e. it has the same value for almost
all realizations of disorder, a part a set of zero probability measure. This
is not in  contradiction with  the fact that the overlap probability
distribution could be not self-averaging. Indeed, it is possible to prove
[5] that the self-averaging of the Lyapunov exponent
only implies that the extrema of the support of $P(q)$ are the same for all
configurations of disorder. Consider the quantity
$$
\Gamma_N(\omega)=
\ln \int_{-1}^{1} \e^{N\, \omega \, q} \, P(q;N) \ \d q
\eqno(4.3)$$
where $P(q;N)$ is the overlap probability distribution of a system of size N,
and in general depends on the realization of disorder.
The Lyapunov exponent is related to the thermodynamic limit
of
$$
{1\over N} \, \Gamma_N(\omega)= {1\over N}
\ln \int_{-1}^{1} e^{N\, \omega \, q} \, P(q;N) \ dq
\eqno(4.4)$$
In the limit of large $N$, we can insert into (4.4) the saddle point estimate
obtaining for small $\omega$
$$
\Gamma(\omega)= \cases{  \omega \, q_{\rm max}
                               \qquad {\rm if}  \ \omega>0 \cr
                               \omega \,  q_{\rm min}
                               \qquad {\rm if}  \    \omega<0 }
\eqno(4.5)$$
In general, $\Gamma(\omega)$ is  a non-linear function of $\omega$.

A reasonable ansatz on the finite $N$ corrections to the asymptotic form of
$P(q)$ for a given disorder realization is
$$
P(q;N) \sim P(q;\infty)+ A \, e^{-S(q) N}.
$$
In this case, the saddle point estimate of (4.4)
gives the Lyapunov exponent as the Legendre transform of
the convex envelope $s(q)$ of $S(q)$:
$$
\Gamma(\omega) = \max_q \ [q \omega - s(q)]
\eqno(4.6)
$$
As consequence of the Oseledec theorem, (4.6) shows that
the envelope $s(q)$ is self-averaging  in systems with
short range interactions. By definition $s(q)=0$ for
$q \in [q_{\rm min},q_{\rm max}]$
and $s(q)>0$ for $q_{\rm max} < q \leq 1$ and $-1 \leq  q <q_{\rm min}$.
Moreover, for large $|\omega|$ the saddle point is given by $q=\pm 1$
so that the asymptotic behavior is
$\Gamma(\omega) \simeq C_{(\pm)}  \pm \omega$
where $C_{(\pm)}$ are constants.

This result could appear of rather limited utility, but it assumes a
great importance when considered as a mark of a replica symmetry breaking.
Indeed,  if $\lim_{N \to \infty} P(q,N)=\delta(q-\avd{q})$, the derivative
$\d\Gamma(\omega) / \d\omega$ at $\omega=0$
does exist and is equal to  $\avd{q}$.
This is the case in one-dimensional systems when $T \neq 0$.
On the other hand,  if $P(q)$ differs from a delta function
it implies a non-differentiable $\Gamma(\omega)$ at $\omega=0$.

No information is lost if $P(q;\infty)$ is a delta function, implying
$q_{\rm min}=q_{\rm max}=\avd{q}$, and $s(q)$ has only one zero at
$q=\avd{q}$.

The theoretical relevance of this result follows from
the possibility of estimating the Lyapunov exponent either by
a direct numerical calculation or by many analytic methods such as
weak disorder expansions [8], cycle expansions of appropriate zeta functions
[9], microcanonical tricks [10-11], and so on.

This approach has been used in Ref.~5 to evaluate the average
overlap
$\avd{q}$ for the model with $p=1/2$ for both zero and finite temperature.
In this reference  $\avd{q}$ was obtained from the numerical derivative of
$\Gamma(\omega)$ at $\omega=0$. Moreover, the zero temperature case was done
by using an {\it ad hoc} trick to extract
the diverging part. Here we report some results on the model with
$0< p\leq 1/2$ for both zero and finite temperature obtained by using
a different method [4] which avoids numerical  estimates
of derivatives.
As we discuss in Appendix 2, the entropy $s(q)$ defined in (4.6) is the
Legendre transform of $(1/ N)\,\avd{\ln {\cal N}_{N}(\omega)}$.
The quantity ${\cal N}_{N}(\omega)$ can be written as the
weighted sum of $\exp(N\omega q)$ over the states with a given value of $q$,
   $${\cal N}_{N}(\omega)= \sum_{q}\, {\cal N}_{N}(q)
                      e^{N\,\omega\,q}.
    \eqno(4.7)$$
For large $N$ the saddle point estimate yields
   $$-s(q)= {1\over N}\,\avd{\ln {\cal N}_{N}(q)}=
           {1\over N}\,\avd{\ln {\cal N}_{N}(\omega)} - \omega\,q
   \eqno(4.8)$$
where $q\equiv q(\omega)$ is the overlap selected by the chosen value of
$\omega$.
All the quantities on the r.h.s. of (4.8) can be computed directly by means
of products of suitable transfer matrices. A brief description of the
method is in Appendix 1.
We report the numerical results for the average overlap
as function of the temperature $T=\beta^{-1}$ at $p=1/2$ in Fig. 3
and as function of $p$ at different temperatures in Fig. 4.
We have also apply our method to the direct
computation of the second derivatives, as one can see in
Fig. 5 which gives the behaviour of the spin glass susceptibility
$$
\chi_{SG}\equiv \avt{(q-\avd{q})^2}=
\left. {\partial^2 \Gamma(\omega) \over \partial \omega^2} \right|_{\omega=0}
$$
as function of the temperature at $p=1/2$.

\section{Conclusions}
We have shown that in one dimensional systems with short range disorder,
ergodicity breaking cannot appear at finite temperature. In generic
situations the overlap between two equilibrium configurations does not vanish
and has a self-averaging value which can be computed by products of random
transfer matrices.

To describe the zero temperature phase transition,
a new one dimensional  model has been introduced. It  is  complex enough
to be a good laboratory for testing ideas which can be useful
 in more realistic situations.
In the weak disorder regime, our model exhibits an overlap probability $P(q)$
 which  is  a smooth function of $q$. It does not self-averages
 in the thermodynamic limit but depends on the disorder configuration.
The $P(q)$ has an analytic  form which can be computed
 in a transparent way
 without using replica trick or other indirect methods.
It is an open problem to find analogous models in two dimensions.

\appendix{}
In this appendix, we show that the overlap probability distribution of a
one dimensional
random system with short range interactions is a self-averaging delta
function at any temperature different from zero.

Consider a one dimensional random Ising model with short range interactions.
Because of the well known Landau theorem on the absence
of phase transitions in one dimension,
 in any realization of the random couplings and/or fields,
 the spin-spin correlation should have the exponential decay
$$
\avt{\sigma_i \, \sigma_{i+r}} - \avt{\sigma_i} \, \avt{\sigma_{i+r}}
\sim e^{-r/\xi}
\eqno(A1)
$$
where $\avt{\cdot}$ denotes the thermal average with the Gibbs measure
$\exp(-\beta \, H)$ for a given disorder realization.
Since in a typical realization of disorder $\xi^{-1}$ is the difference of
the  Lyapunov exponent of the product of random transfer matrices related  to
the model, (A1) also follows from
 the Perron theorem  assuring that  there is no degeneracy of the
eigenvalues in the spectrum of matrices with positive non-zero elements.
 The value of $\xi$  can depend on the sites  $i$,  $i+r$
 and on the disorder realization. Nevertheless, (A1)
 implies that the  $\sigma_i$
 are (exponentially) independent  random variables  with respect
 to the Gibbs measure.

The overlap between two spin configurations $\sigma^{1}$ and
$\sigma^{2}$  is
$$
q_{12} \equiv{1\over N} \sum_{i=1}^N \, \avt{\sigma^{1}_i}  \,
            \avt{\sigma^{2}_i}.
$$
As the $\sigma_i$ are independent, we can apply the law of large numbers
 so that in the thermodynamic limit with probability one
 all spin configurations related to a disorder realization
 have the same overlap, i.e.
$\lim_{N \to \infty} q_{12} \to \avd{q}$ and the overlap probability
distribution $P(q)$ converges towards a delta function.

We should still prove that $\avd{q}$ has the same value for all disorder
 realizations. This can be seen by means of a very general result:
the  support of the overlap probability distribution $P(q)$ is self-averaging.
Consequently, if $P(q)$ is a delta function for a disorder
realization, it should be a delta function centered on the same value
$\avd{q}$, for all the other realizations.
In conclusion, for one dimensional systems with short range interactions
one has in the thermodynamic limit for any non-zero temperature
$$
P(q)=\delta(q-\avd{q}).
$$
for almost all disorder realizations,
a part a set of zero probability measure.
\vfill\eject

\appendix{}
In this appendix we briefly review the numerical method used in Sect.~4 to
evaluate the average overlap probability distribution. We only describe the
finite temperature method. We shall denote by
$\bbox{S}_i\equiv (\sigma^{1}_i, \sigma^{2}_i)$ the pair of spin at the
same site in the two replicas. The average of the logarithm of
${\cal N}_{N}(\omega)$, equation (4.1),  can be obtained from
the vector $\bbox{N}_i(\bbox{S}_{i+1})$ which satisfy the recursion relation
   $$\bbox{N}_i(\bbox{S}_{i+1})= \sum_{\bbox{S}_i}
        \exp(-\beta \epsilon^{1}_i - \beta \epsilon^{2}_i
               + \omega \sigma^{1}_i \sigma^{2}_i)\,
        \bbox{N}'_{i-1}(\bbox{S}_i)
   \eqno(B.1)$$
where $\epsilon_i^{\alpha}= -[\sigma^{\alpha}_{i+1} + h_i]\sigma^{\alpha}_i$
is the single spin energy in the site $i$ of the replica $\alpha$, and
   $$\sum_{\bbox{S}_{i+1}} \bbox{N}_i'(\bbox{S}_{i+1})= 1.
    \eqno(B.2)$$
{}From (4.1) follows the initial condition
   $$ \bbox{N}_1(\bbox{S}_{2})= \sum_{\bbox{S}_1}
        \exp(-\beta \epsilon^{1}_1 - \beta \epsilon^{2}_1
               + \omega \sigma^{1}_1 \sigma^{2}_1).
    \eqno(B.3)$$

By denoting with $n_i$ the sum of the elements of
$\bbox{N}_i(\bbox{S}_{i+1})$, the logarithm of ${\cal N}_{N}(\omega)$
averaged over the disorder is
   $$ {1\over N}\avd{\ln {\cal N}_{N}(\omega)}=
         \lim_{N\to\infty} {1\over N}\,\sum_{i=1}^{N} \ln n_i
    \eqno(B.4)$$
i.e. the Lyapunov exponent of the product.

The value of $q$ can be obtained directly from the derivative
of the vector $\bbox{N}_i(\bbox{S}_{i+1})$ with respect to $\omega$.
The derivative with respect to $\omega$
leads to the vector $\bbox{O}_i(\bbox{S}_{i+1})$ obeying the recursion relation
   $$\bbox{O}_i(\bbox{S}_{i+1})=
    \sum_{\bbox{S}_{i}}\,\left[
      \bbox{O}'_{i-1}(\bbox{S}_{i}) +
         \sigma^{1}\sigma^{2} \bbox{N}'_{i-1}(\bbox{S}_{i})
                           \right]\,
      e^{\omega \sigma^{1}_i\sigma^{2}_i}\,
     \prod_{\alpha=1,2} e^{-\beta \epsilon^{\alpha}_i}
    \eqno(B.5)$$
with initial condition
   $$\bbox{O}_1(\bbox{S}_{2})= \sum_{\bbox{S}_1}
             \sigma^{1}_1 \sigma^{2}_1\,
        \exp(-\beta \epsilon^{1}_1 - \beta \epsilon^{2}_1
               + \omega \sigma^{1}_1 \sigma^{2}_1).
    \eqno(B.6)$$
At each step the vector $\bbox{O}_i$ is rescaled as
   $$\bbox{O}'_{i}(\bbox{S}_{i+1})=
              \left[
     \bbox{O}_{i}(\bbox{S}_{i+1})
       - o_i\, \bbox{N}'_{i}(\bbox{S}_{i+1})
              \right] / n_i
    \eqno(B.7)$$
where $n_i$ and $o_i$ are the sum of the elements of
$\bbox{N}_{i}(\bbox{S}_{i+1})$ and $\bbox{O}_{i}(\bbox{S}_{i+1})$,
respectively. The value of $q=q(\omega)$ is then obtained from the average of
$o_i$
   $$q= \lim_{N\to\infty} \frac{1}{N}\,\sum_{i} \frac{o_i}{n_i}.
    \eqno(B.8)$$

In the limit of zero temperature one has to extract the diverging part in
(4.8). This is achieved by means of a Legendre transform to express all the
quantities in terms of the average energy per spin
$\epsilon\equiv\epsilon(\omega,\beta)$. The value of $\epsilon$ for given
$\beta$ and $\omega$ can be obtained by introducing a vector of the
derivative of $\bbox{N}_i(\bbox{S}_{i+1})$ with respect to $\omega$, ad
proceeding as done for $q$.

\references

\numrefbk{[1]}{M\'ezard M., Parisi G. and Virasoro M.A.,}{Spin Glass Theory
       and Beyond}{(World Scientific, Singapore 1988);}

\numrefjl{[2]}{Sherrington D. and Kirkpatrick S.,}
       {Phys. Rev. Lett.}{32}{(1975) 792;}
\numrefjl{   }{Kirkpatrick S. and Sherrington D.,}
       {Phys. Rev. B}{17}{(1978) 4384.}

\numrefjl{[3]}{Derrida B.,}
       {Phys. Rev. B}{24}{(1981) 2613}

\numrefjl{[4]}{Crisanti A., Paladin G., Serva M. and Vulpiani A.,}
       {Phys. Rev. Lett.}{}{(1993) submitted}

\numrefjl{[5]}{Crisanti A., Paladin G., Serva M. and Vulpiani A.,}
       {Phys. Rev. Lett.}{}{(1993) submitted}

\numrefjl{[6]} {Franz S., Parisi G. and Virasoro M.,}
       {J. de Physique I}{2}{(1992) 1869}

\numrefjl{[7]}{Oseledec V.I.,}{Trans. Mosc. Math Soc.}{19}{(1968) 197.}

\numrefjl{[8]}{Derrida B., Mecheri K., and Pichard J.L.,}
       {J. de Physique}{48}{(1987) 733}

\numrefjl{[9]}{Mainieri R.,}
       {Phys. Rev. Lett.}{68}{(1992) 1965}

\numrefjl{[10]}{Deutsch J. and Paladin G.,}
       {Phys. Rev. Lett.}{62}{(1988) 695}

\numrefjl{[11]}{Serva M. and Paladin G.,}
       {Phys. Rev. Lett.}{70}{(1993) 105}

\figures
\figcaption{Overlap probability $P_2(q;x)$ for $x=0.1$ (a), $x=0.2$ (b) and
            $x=0.5$ (c). The dashed line is the average overlap probability
            $\avd{P_2}(q)=  \int_{0}^{1} \, dx \, P_2(q;x)$.}

\figcaption{Overlap probability distributions:  $P_{3}(q;x_1,x_2)$
            with impurities at $x_1=1/3$ and $x_2=1/3$ (full line),
            average probability $\avd{P_3}(q)$ (dashed line) and gaussian
            distribution with mean value $1/3$ and variance $2/27$ (dot
            dashed line). }

\figcaption{Numerical results for the average overlap $\avd{q}$
            as function of the temperature $T=\beta^{-1}$ in the disordered
            system with $p=0.5$ (a), $p=0.1$ (b) and $p=0.01$ (c).}

\figcaption{Numerical results for the average overlap $\avd{q}$
            as function of the disorder concentration $p$ at temperatures
            $T=0$ (a), $T=0.5$ (b), $T=1$ (c) and $T=2$ (d). }

\figcaption{Spin glass susceptibility $\chi_{SG}\equiv \avt{(q-\avd{q})^2}$
            as function of the temperature in the disordered system with
            $p=0.5$ (a) and $p=0.1$ (b).}

\bye